\documentstyle[a4wide,12pt]{article}
\title{A discussion on a possibility to interpret\\ quantum mechanics in
terms of general relativity}
\author{Vu B Ho\\Department of Physics\\Monash University\\
Clayton Victoria 3168\\Australia}
\date{}
\begin{document}
\maketitle

\begin{abstract}
It is shown that, with some reasonable assumptions, the theory of general
relativity can be made compatible with quantum mechanics by using the
field equations of general relativity to construct a Robertson-Walker
metric for a quantum particle so that the line element of the particle can
be transformed entirely to that of the Minkowski spacetime, which is
assumed by a quantum observer, and the spacetime dynamics of the particle
described by a Minkowski observer takes the form of quantum mechanics.
Spacetime structure of a quantum particle may have either positive
or negative curvature. However, in order to be describable using the
familiar framework of quantum mechanics, the spacetime structure of a
quantum particle must be "quantised" by an introduction of the imaginary
number $i$. If a particle has a positive curvature then the quantisation
is equivalent to turning the pseudo-Riemannian spacetime of the particle
into a Riemannian spacetime. This means that it is assumed the particle is
capable of measuring its temporal distance like its spatial distances. On
the other hand, when a particle has a negative curvature and a negative
energy density then quantising the spacetime structure of the particle is
equivalent to viewing the particle as if it had a positive curvature and
a positive energy density.
\end{abstract}
\newpage

It has been considered that general relativity may not be compatible with
the quantum theory because the former is formulated in terms of curved
spacetimes while the later is based on the view of an observer who uses the
Minkowski spacetime and describes the quantum dynamics of a particle in
terms of a Hilbert space of physical states. However, it can be seen that
the two descriptions may reconcile if a curved spacetime of a quantum
particle can be transformed to the Minkowski spacetime so that a Minkowski
formulation of the spacetime dynamics of the particle is that of the quantum
theory. The following is a discussion of such a possibility.

Consider first the following situation \cite{Ho}. If the dynamics of a
particle is assumed to be described by the field equations of general
relativity
\begin{equation}
R_{\mu\nu}-\frac{1}{2}g_{\mu\nu}R+g_{\mu\nu}\Lambda=\kappa T_{\mu\nu}
\end{equation}
then with $\Lambda=0$ and a centrally symmetric spacetime metric
\cite{Land}
\begin{equation}
ds^2=e^\mu dt^2-e^\nu dr^2-r^2(d\theta^2+\sin^2\theta d\phi^2)
\end{equation}
and an energy-momentum tensor of the form
\begin{equation}
T_\mu^\nu = \left(\begin{array}{cccc}
-\frac{\alpha\beta}{\kappa}\frac{e^{-\beta r}}{r^2} & 0 & 0 & 0\\
0 & -\frac{\alpha\beta}{\kappa}\frac{e^{-\beta r}}{r^2} & 0 & 0\\
0 & 0 & \frac{\alpha\beta^2}{2\kappa}\frac{e^{-\beta r}}{r} & 0\\
0 & 0 & 0 & \frac{\alpha\beta^2}{2\kappa}\frac{e^{-\beta r}}{r}
\end{array}\right),
\end{equation}
the field equations of general relativity will admit as an exact solution
the following line element of Yukawa potential form
\begin{equation}
e^{-\nu}=1-\alpha\frac{e^{-\beta r}}{r}+\frac{Q}{r},
\end{equation}
where the term $Q/r$ could be interpreted as the Coulomb repulsion force of
a proton. Because the physical system is expressed in terms of
curved spacetime of the particle, whether this solution is physically
acceptable, although it is mathematically acceptable, could not be justified
under the view of an observer who uses the Minkowski spacetime, unless the
above curved spacetime of the particle can be transformed entirely to a
manifestly Minkowski metric. However, the above form of the energy-momentum
tensor seems to reveal a possibility that at the quantum level the energy
density may vary only as an inverse square of distance and the pressure may
be ignored compared to the energy density. With these observations, let us
now consider a general relativistic spacetime model for a quantum particle
using the Robertson-Walker metric \cite{Narl,Peeb}
\begin{equation}
ds^2=c^2dt^2-S^2(t)\left(\frac{dr^2}{1-kr^2}+r^2(d\theta^2 + \sin^2\theta
d\phi^2)\right)
\end{equation}
with the energy-momentum tensor $T_{\mu\nu}$ of the form
\begin{equation}
T_\mu^\nu= \left(\begin{array}{cccc}
\frac{A}{S^2} & 0 & 0 & 0\\0 & 0 & 0 & 0\\0 & 0 & 0 & 0\\0 & 0 & 0 & 0
\end{array}\right).
\end{equation}
The field equations of general relativity then reduce to the system
\begin{eqnarray}
\frac{\dot{S}^2}{S^2}+\frac{kc^2}{S^2}-\frac{\Lambda c^2}{3}&=&\frac{\kappa
c^2}{3}\frac{A}{S^2}\\
2\frac{\ddot{S}}{S}+\frac{\dot{S}^2}{S^2}+\frac{kc^2}{S^2}-\Lambda c^2&=&0.
\end{eqnarray}
This system of equations has a static solution
\begin{equation}
S_0^2=\frac{kc^4}{4\pi G\epsilon}
\end{equation}
which is similar to the Einstein static model with an energy density $\epsilon$
which may be very large \cite{Eins}. However, in this case, the possibility of
negative energy density should not be ruled out because at the quantum level
a particle may have a curved spacetime with negative curvature as will be
discussed in the following. Since we are discussing curved spacetimes at the
quantum level, the quantity $\Lambda = k/S_0^2$ will change drastically for a
small fraction of variation of $S_0$. Hence, the quantity $\Lambda$ should
also be considered as an inverse square function of $S$, i.e.
$\Lambda = B/S^2$, where $B$ is constant (actually the constant $B$ could be
set to zero for almost all arguments that follow from here). The above system
of field equations is then modified to the system of equations
\begin{eqnarray}
\frac{\dot{S}^2}{S^2}+\frac{kc^2}{S^2}-\frac{c^2}{3}\frac{B}{S^2}
&=&\frac{\kappa c^2}{3}\frac{A}{S^2}\\
2\frac{\ddot{S}}{S}+\frac{\dot{S}^2}{S^2}+\frac{kc^2}{S^2}-c^2\frac{B}{S^2}
&=&0.
\end{eqnarray}
This system of equations has a solution of the form
\begin{equation}
S=act, \ \ \ \ \ \ \mbox{where} \ \ \ \ \ a=\sqrt{\frac{\kappa A}{2}-k}
\end{equation}

Let us first consider the case $k=1$. It is seen that in this case a real
solution requires spacetime of a quantum particle to have a very large
positive energy density in its own reference frame. However, a particle
with this large positive energy density can not be viewed by an observer
who uses the Minkowski spacetime because the curved spacetime of the particle
in this case can not be transformed to the Minkowski spacetime of a quantum
observer. Furthermore, such a large energy density is not appropriate
for the surrounding spacetime of quantum particles like protons and
neutrons. Therefore, if we assume a reasonable value for the energy density
so that $\kappa A/2\ll 1$, then we are forced to "quantize" the
spacetime structure of the particle by introducing the imaginary number $i$
and let $a\approx i$ or $S\approx ict$. Actually, this kind of quantisation
turns the pseudo-Riemannian curved spacetime of the particle into a
Riemannian spacetime. This means that we assume the particle to be able to
measure its temporal distance in exactly the same way as its spatial
distances. The quantisation is realisable only when the curved spacetime of
the particle can be viewed in the Minkowski spacetime. This is in fact the
case for if we apply the coordinate transformations \cite{Narl}
\begin{equation}
iR=ctr, \ \ \ \ \ cT=ct\sqrt{1-r^2}
\end{equation}
then, as can be verified, these coordinate transformations reduce the
Robertson-Walker metric of the quantum particle to a manifestly Minkowski
metric of the form
\begin{equation}
ds^2=c^2dT^2-dR^2-R^2(d\theta^2+\sin^2\theta d\phi^2).
\end{equation}
It is seen that the dynamics of spacetime structure of the quantum particle
can now be investigated by an observer whose uses a Minkowski metric. The
investigation can be carried out by writing the quantity $S$ in terms of
the coordinates $(R,cT)$ in the form of an action integral
\begin{equation}
S=-i\sqrt{c^2T^2-R^2}=-i\int ds=-ic\int\sqrt{1-\frac{v^2}{c^2}}dT
\end{equation}
where $ds$ is the usual Minkowski spacetime interval and $v=R/T$. In
addition, if we define a new quantity $\Psi$ by the relation $S=K\ln\Psi$,
then we have
\begin{equation}
\Psi= e^{\frac{i}{K}\int ds}.
\end{equation}
With this form, the familiar quantum mechanics can be recovered by applying
the Feymann path integral method \cite{Feyn}. However, since in Minkowski
spacetime the quantity $S$ has an action integral form, we can construct a
quantum mechanics by following Schr\"{o}dinger's method \cite{Your} as in
his original derivation of the wave equation of quantum mechanics by
observing that the quantity $S$ satisfies the relation
\begin{equation}
-\frac{1}{c^2}\left(\frac{\partial S}{\partial T}\right)^2 + \left(
\frac{\partial S}{\partial R})\right)^2-1=0.
\end{equation}
The quantity $\Psi$ then satisfies the relation
\begin{equation}
\frac{1}{c^2}\left(\frac{\partial\Psi}{\partial T}\right)^2 - \left(
\frac{\partial \Psi}{\partial R}\right)^2 - \frac{1}{K^2}\Psi^2=0.
\end{equation}
Since $\partial\Psi/\partial R=\nabla\Psi .\partial{\bf R}/\partial
R=|\nabla\Psi|\cos\alpha$, using the variational principle, after averaging
the above equation with $<\cos^2\alpha>=1/2$, we obtain a Klein-Gordon-like
wave equation \cite{Mess}
\begin{equation}
-\frac{1}{c_a^2}\frac{\partial^2 \Psi}{\partial T^2}+\nabla^2\Psi -
\frac{1}{K_a^2}\Psi=0,
\end{equation}
where $c_a=c/\sqrt{2}$ and $K_a=K/\sqrt{2}$. If we compare this equation
with the Klein-Gordon equation in quantum mechanics, then we see that this
equation describes a quantum dynamics of a particle with an average
velocity $c_a$, rather than that of light $c$. This may be a reason why
the force carriers in strong and weak interactions have mass. The
comparison also gives $K_a=\hbar/mc_a$. Here, perhaps, the most important
point that should be emphasised is that the Minkowski coordinates in this
case depend entirely on the metric structure of a quantum particle. So
observers who use the Minkowski spacetime can not perform measurements of
physical observables of the particle by their own choices of gauges of space
and time. This may be the reason for unpredictable behaviours of quantum
particles in the Minkowski quantum mechanics.

Now let us consider the case $k=-1$. Similar to the previous discussions,
however, in this case we assume $a\approx 1$ or $S\approx ct$. The coordinate
transformations of the form \cite{Narl}
\begin{equation}
R=ctr, \ \ \ \ \ cT=ct\sqrt{1+r^2}
\end{equation}
also reduce the Robertson-Walker metric of a quantum particle to that of the
Minkowski spacetime. The quantity $S$ written in terms of the
coordinates  $(R,cT)$ takes the form
\begin{equation}
S=\sqrt{c^2T^2-R^2}=\int ds=c\int\sqrt{1-\frac{v^2}{c^2}}dT
\end{equation}
and satisfies the relation
\begin{equation}
-\frac{1}{c^2}\left(\frac{\partial S}{\partial T}\right)^2 + \left(
\frac{\partial S}{\partial R}\right)^2+1=0.
\end{equation}
The quantity $\Psi$ now becomes
\begin{equation}
\Psi=e^{\frac{1}{K}\int ds}
\end{equation}
and satisfies the equation
\begin{equation}
-\frac{1}{c_a^2}\frac{\partial^2 \Psi}{\partial T^2}+\nabla^2\Psi +
\frac{1}{K_a^2}\Psi=0,
\end{equation}
This equation differs from the Klein-Gordon equation by the plus sign
before the last term. This results from the fact that the quantum particle
in this case has negative curvature. This kind of structure of a particle
is not assumed in quantum mechanics. However, it is seen that if the energy
density in this case is negative then every thing up to this stage
is real and compatible with the usual relativistic description with timelike
intervals. If we want to "quantise" spacetime structure by turning
the above equation into the Klein-Gordon equation so that we can
describe the quantum dynamics of the particle using the familiar quantum
mechanics then we can let $K\rightarrow iK$. This process of quantisation is
equivalent to turning spacetime structure of a particle with negative
curvature into that with positive curvature and specifying a positive energy
density for the particle in the Minkowski spacetime.

Finally consider the case $k=0$. In this case a real solution is obtained for
any positive energy density, $S=act$. If we apply the coordinate
transformations
\begin{equation}
R=actr, \ \ \ \ \ \ cT=ct
\end{equation}
then $dR=actdr+acrdT$. It is seen that when the term $acrdT\ll 1$, the
spacetime structure of a particle can be reduced to that of the Minkowski
spacetime. A particle with large energy density, i.e. $a$ large, can only
appear to a Minkowski observer for a short time $dT$. On the other hand, a
particle with low energy density can exist with respect to a Minkowski observer
for a long period of time $dT$.

As a conclusion we make some remarks about the energy-momentum conservation
laws in general relativity. For simple models that have been discussed the only
solution $S=S_0$ satisfies strictly the conservation laws required by the
general theory. However, when the theory is applied into quantum physics the
conservation laws should not be expected to satisfy at the quantum level, and
the uncertainty principle would allow any such violation. This can also be seen
by spacetime coordinate relationships discussed above. Furthermore, it seems
that at the quantum level concepts like positiveness of energy density also
become relative and coordinate-dependent, and this may affect the foundations
of physics of a Minkowski observer. These sophisticated problems require
further investigations.

\section*{Acknowledgements}
I would like to acknowledge the financial support of an APA Research Award.

\end{document}